\documentclass[conference]{IEEEtran}
\pdfoutput=1
\usepackage[colorlinks]{hyperref}

\usepackage{amsmath}
\usepackage{amsthm}
\usepackage{amssymb}
\usepackage{bm}
\usepackage{xspace}
\usepackage{xcolor}
\usepackage{graphicx}
\usepackage{url}
\usepackage{framed}
\usepackage{float}
\usepackage{rotating}
\usepackage{verbatim}
\usepackage{listings}
\usepackage{lscape}

\usepackage{pgfplots}
\usepackage{pgf}
\usepackage{tikz}
\usetikzlibrary{arrows,shapes.misc,chains,scopes}
\usepackage{pgfplotstable}
\usepackage{booktabs}
\usepackage{colortbl}

\newcommand{\executeiffilenewer}[3]{%
\ifnum\pdfstrcmp{\pdffilemoddate{#1}}%
{\pdffilemoddate{#2}}>0%
{\immediate\write18{#3}}\fi%
}

\graphicspath{{figures/}}

\theoremstyle{plain}
\newtheorem{proposition}{Proposition}

\newtheorem{remark}{Remark}

\newcounter{algocount}
\newcounter{examplecount}

\newcommand{\setu}{\ensuremath{\mathcal{U}}\xspace}

\newcommand{\setx}{\ensuremath{\mathcal{X}}\xspace}

\newcommand{\bmm}{\begin{matrix}}
\newcommand{\emm}{\end{matrix}}
\newcommand{\bpm}{\begin{pmatrix}}
\newcommand{\epm}{\end{pmatrix}}

\newcommand{\bsbm}{\left[\begin{smallmatrix}}
\newcommand{\esbm}{\end{smallmatrix}\right]}

\newcommand{\bbm}{\begin{bmatrix}}
\newcommand{\ebm}{\end{bmatrix}}

\DeclareMathOperator*{\argmin}{argmin}

\DeclareMathOperator{\expop}{\mathbb{E}}
\DeclareMathOperator{\entop}{\mathbb{H}}

\DeclareMathOperator{\kl}{\mathbb{D}}

\usepackage{cite}
\usepackage{printlen}

\newcommand\numberthis{\addtocounter{equation}{1}\tag{\theequation}}
\title{Fixed-to-Variable Length Resolution Coding\\
for Target Distributions}

\author{\IEEEauthorblockN{Georg B\"ocherer and Rana Ali Amjad}
\IEEEauthorblockA{Institute for Communications Engineering\\Technische Universit\"at M\"unchen, Germany\\
Email: \texttt{georg.boecherer@tum.de,raa2463@gmail.com}}
\thanks{This work was supported by the German Ministry of Education and Research in the framework of an Alexander von Humboldt Professorship.}
}

\newcommand{\overleq}[1]{\overset{(\mathrm{#1})}{\leq}}
\newcommand{\overgeq}[1]{\overset{(\mathrm{#1})}{\geq}}
\newcommand{\lmax}{{\ell_{\max}}}

\DeclareMathOperator{\supp}{supp}

\begin{document}

\maketitle

\begin{abstract}
The number of random bits required to approximate a target distribution in terms of un-normalized informational divergence is considered. It is shown that for a variable-to-variable length encoder, this number is lower bounded by the entropy of the target distribution. A fixed-to-variable length encoder is constructed using $M$-type quantization and Tunstall coding. It is shown that the encoder achieves in the limit an un-normalized informational divergence of zero with the number of random bits per generated symbol equal to the entropy of the target distribution. Numerical results show that the proposed encoder significantly outperforms the optimal block-to-block encoder in the finite length regime. 
\end{abstract}

\section{Introduction}

Given is a target distribution $P_Y$. We ask the following question.
\begin{quote}
What is the minimum number of random bits that we need to generate symbols that appear to be distributed according to $P_Y$?
\end{quote}
This minimum number is called the \emph{resolvability} of the target distribution and codes that achieve this minimum are called \emph{resolution codes}. This question is a special case of the problem of channel resolvability, namely when the channel input is equal to the channel output. In  \cite{han1993approximation}, this special case is called the identity channel. For channel resolvability, three measures of resemblance have been considered in the literature, namely the normalized informational divergence, the un-normalized informational divergence, and the variational distance. Wyner \cite{wyner1975} has discussed resolvability for \emph{Discrete Memoryless Channels} (DMC) and product target distributions. Using normalized informational divergence, he showed that the minimum number of bits required for this task is the mutual information of channel input and channel output. In \cite{han1993approximation}, Han and Verdu have shown similar results for information stable distributions using both variational distance and normalized informational divergence. It should be pointed out that neither criterion (i.e. normalized informational divergence and variational distance) is stronger than the other. A stronger result using un-normalized informational divergence has been shown in \cite{jiehou2013} for DMCs and product distributions, implying both the results in \cite{wyner1975} and \cite{han1993approximation}. The analysis presented in \cite{wyner1975,han1993approximation,jiehou2013} is based on random coding arguments and only the \emph{existence} of resolvability achieving codes is shown. No discussion has been presented on the construction of practical encoders. Moreover, the problem of channel resolvability has only been addressed in the context of block-to-block encoders.

For \emph{normalized} informational divergence and the identity channel, distribution matching codes can be used. Variable length codes that achieve resolvability have been developed in \cite{bocherer2011matching},\cite[Sec.~3.2]{bocherer2012capacity} and \cite{amjad2013fixed}. These codes are one-to-one, i.e., the input can be decoded from the output, a property that is normally not characteristic for resolution codes. Bloch \emph{et al} have discussed in \cite{bloch2012} the use of polar codes to construct resolution codes for binary input symmetric DMCs using un-normalized informational divergence.

In this work, we consider the problem of resolution coding for target distributions using variable length encoders. We use un-normalized informational divergence as our criterion for approximation. The resolution rate of a variable length encoder is defined in Sec.~\ref{sec:resolvability} and in Sec.~\ref{sec:hv}, we relate our definition to the definitions in \cite{han1993approximation}. We lower bound achievable resolution rates by the target distribution entropy in Sec.~\ref{sec:converse}. We then propose in Sec.~\ref{sec:f2v} a  fixed-to-variable length encoder using Tunstall coding \cite[Sec.~2.10]{kramer2012information} and $M$-type quantization \cite{bocherer2013optimala}. In Sec.~\ref{sec:proof}, we prove that our scheme achieves in the limit the lower bound. Finally, in Sec.~\ref{sec:numerical}, we present numerical results that show that our fixed-to-variable length scheme significantly outperforms the optimal block-to-block scheme in the finite length regime.

\section{Variable Length Resolution Coding}
\label{sec:resolvability}

\subsection{Target Distribution}
Consider a \emph{Discrete Memoryless Source} (DMS) $P_Y$ that generates a sequence $Y_1,Y_2,\dotsc$ where the $Y_i$ are iid according to $P_Y$ and where $Y_i$ takes values in a finite set $\mathcal{Y}$. We define $D:=|\mathcal{Y}|$, i.e., the DMS $P_Y$ generates $D$-ary strings. Denote by $\mathcal{X}$ the set of paths from the root to the leaves of a complete $D$-ary tree. A $D$-ary tree is complete if every right-infinite $D$-ary sequence starts with a path from $\setx$. A path $x$ can be written as $x=x_1x_2\dotsb x_{\ell(x)}$ with $x_i\in\mathcal{Y}$, where $\ell(x)$ denotes the length of path $x$. By using the distribution $P_Y$ as a branching distribution in the tree, we define a distribution $P_Y^\mathcal{X}$ over $\mathcal{X}$ as follows \cite[p.~23]{bocherer2013rooted}.
\begin{align*}
P_Y^\mathcal{X}(x):=\prod_{i=1}^{\ell(x)}P_Y(x_i).
\end{align*}
If $\mathcal{X}=\mathcal{Y}^n$, then $P_Y^\mathcal{X}=P_Y^n$, i.e., the product distribution. Given a complete $D$-ary tree $\mathcal{X}$, the problem of generating a sequence that resembles the output of the DMS $P_Y$ can be solved by generating a sequence that resembles the output of the DMS $P_Y^\mathcal{X}$, see \cite[Prop.~7]{bocherer2013rooted}.

\subsection{Variable Length Encoder}
\label{sec:encoder}

\begin{figure}
		\tikzstyle{transition}=[circle,draw=black!50,fill=black!20,thick,inner sep=0pt,minimum size=3mm]
		\tikzstyle{leaf}=[circle,draw=green!50!black,fill=green!20,thick,inner sep=0pt,minimum size=4mm]
		\tikzstyle{emptyleaf}=[circle,draw=red!50,fill=red!20,thick,inner sep=0pt,minimum size=4mm]
		\tikzstyle{label0} = [above]
		\tikzstyle{label1} = [below]

\begin{tikzpicture}[node distance=0cm, grow=right, scale=0.9,level 1/.style={sibling distance=2cm,level distance=1cm},level 2/.style={sibling distance=1.8cm,level distance=1cm},level 3/.style={sibling distance=1cm,level distance=1cm},unit/.style={rectangle,fill=black!5,draw,thick,minimum size = 0.75cm, text width = 1.5cm, text centered}]
\node[transition] at (6,0) {}
	child {node[transition] {}
		child {node[transition] {}
			child {node[leaf](111) {}
				edge from parent[->]
				node[label1]{$1$}
			}
			child {node[leaf](110){}
				edge from parent[->]
				node[label0]{$0$}
			}
			edge from parent[->]
			node[label1]{$1$}
		}
		child {node[transition] {}
			child {node[leaf](101) {}
				edge from parent[->]
				node[label1]{$1$}
			}
			child {node[leaf](100) {}
				edge from parent[->]
				node[label0]{$0$}
			}
			edge from parent[->]
			node[label0]{$0$}
		}
	edge from parent[->]
	node[label1]{$1$}
	}
	child {node[leaf](0) {}
	edge from parent[->]
	node[label0]{$0$}
	};
\node [right=of 110](0110){$110$};
\node [right=of 0](00){$0$};
\node [right=of 101](0101){$101$};
\node [right=of 100](0100){$100$};
\node [right=of 111](0111){$111$};

\node[transition] at (14.5,-1) {}[grow=left,level 1/.style={sibling distance=1.2cm,level distance=1cm},level 2/.style={sibling distance=1cm,level distance=1cm}]
	child {node[transition] {}
		child {node[leaf](aa) {}
				edge from parent[->]
				node[label0]{$a$}
		}
		child {node[emptyleaf](ab) {}
				edge from parent[->]
				node[label0]{$b$}
		}
		child {node[leaf](ac) {}
				edge from parent[->]
				node[label0]{$c$}
		}
				edge from parent[->]
				node[label0]{$a$}
	}
	child {node[emptyleaf](b) {}
				edge from parent[->]
				node[label0]{$b$}
	}
	child {node[leaf](c) {}
				edge from parent[->]
				node[label0]{$c$}
	};
\node [left=of aa](AA){$aa$};
\node [left=of ab](AB){$ab$};
\node [left=of ac](AC){$ac$};
\node [left=of b](B){$b$};
\node [left=of c](C){$c$};

\draw[|->] (0110) -- (C);
\draw[|->] (00) -- (AA);
\draw[|->] (0101) -- (AC);
\draw[|->] (0100) -- (AA);
\draw[|->] (0111) -- (C);

\end{tikzpicture}
\caption{A variable length encoder with a complete dictionary $\setu=\{0,100,101,110,111\}$ and a complete codebook $\setx=\{aa,ab,ac,b,c\}$. The mapping is many-to-one and not onto.}
\label{fig:encoder}
\end{figure}
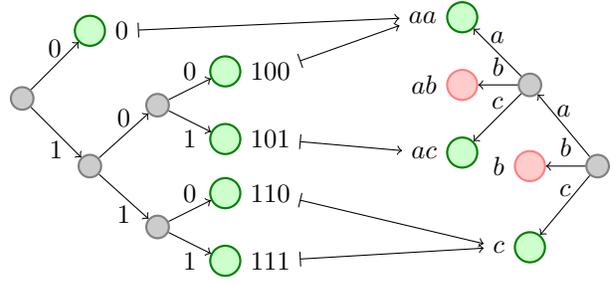

A variable length encoder consists of a complete binary dictionary $\mathcal{U}$, a complete codebook $\mathcal{X}$ (a set is complete if it can be represented by a complete tree), and a \emph{deterministic} mapping $f\colon\mathcal{U}\to\mathcal{X}$. There is no further restriction on $f$, i.e., $f$ may map no, one, or more than one word from $\mathcal{U}$ to a specific codeword from $\mathcal{X}$. We give an example in Fig.~\ref{fig:encoder}.

The encoder parses independent and uniformly distributed bits at its input by its dictionary $\mathcal{U}$. This generates a random variable $U$, which takes values in $\mathcal{U}$ according to $P_U$ with 
\begin{align*}
P_U(u)=2^{-\ell(u)}, \qquad\forall u\in\mathcal{U}. 
\end{align*}
The encoder maps $U$ to a codeword $X=f(U)\in\mathcal{X}$ and the generated distribution is $P_X$. Define
\begin{align*}
\lmax:=\max_{u\in\mathcal{U}}\ell(u).
\end{align*}
The distribution $P_X$ is $2^{\lmax}$-type, i.e., each probability is of the form $k/2^\lmax$ where $k$ is a non-negative integer. The described setting comprises fixed-to-fixed, variable-to-fixed, fixed-to-variable, and variable-to-variable length encoders.
\begin{remark}
If the mapping $f$ is many-to-one, it may generate non-dyadic $2^{\lmax}$-type distributions over $\mathcal{X}$. This is an important difference to channel matching \cite{bocherer2011matching},\cite[Sec.~3.2]{bocherer2012capacity}, where we can only generate dyadic distributions because of a one-to-one mapping constraint for decodability. 
\end{remark}
\subsection{Resolution Rate, Entropy Rate, and Resolvability}
\label{sec:def}
We define the \emph{resolution rate} $R$ of a variable length encoder as the average input length divided by the average output length, i.e.,
\begin{align*}
R(\mathcal{U},\mathcal{X},f):=\frac{\expop[\ell(U)]}{\expop[\ell(X)]}.
\end{align*}
For notational convenience, we also write $R$ if the considered encoder is clear from the context. A resolution rate $R$ is called \emph{achievable} if there exists a family of encoders $(\mathcal{U}_k,\mathcal{X}_k,f_k)$ with resolution rate $R_k$ such that 
\begin{align}
\kl(P_{X_k}\Vert P_Y^{\mathcal{X}_k})\overset{k\to\infty}{\to}&0\label{eq:kllimit}\\
R_k\overset{k\to\infty}{\to}&R.\label{eq:ratelimit}
\end{align}
For the rest of this paper, we will omit the index $k$. For example, $\kl(P_X\Vert P_Y^\mathcal{X})\to 0$ means \eqref{eq:kllimit}. The minimum of all achievable resolution rates is called the \emph{resolvability of} $P_Y$ and is denoted by $S(P_Y)$.
The \emph{entropy rate} of a variable length encoder is \cite[Sec.~4.2]{bocherer2012capacity}
\begin{align*}
\bar{R}(\mathcal{U},\mathcal{X},f):=\frac{\entop(P_X)}{\expop[\ell(X)]}.
\end{align*}
An \emph{achievable entropy rate} is defined accordingly. This is an extension of the definition given in \cite[Sec.~2]{han1993approximation} to variable length encoders. The minimum achievable entropy rate is called \emph{mean resolvability} of $P_Y$ and is denoted by $\bar{S}(P_Y)$.

The expected input length is bounded as
\begin{align*}
\expop[\ell(U)]&=\sum_{u\in\mathcal{U}}2^{-\ell(u)}\ell(u)\\
&=\entop(U)\\
&\geq\entop[f(U)]\\
&=\entop(P_X).\numberthis\label{eq:elu}
\end{align*}
Note that \eqref{eq:elu} is the converse \cite[Theo.~5.11.1]{cover2006elements} for exact random number generation \cite{knuth1976complexity}. Using \eqref{eq:elu}, for any encoder, resolution rate and entropy rate relate as 
\begin{align}
R(\mathcal{U},\mathcal{X},f)\geq\bar{R}(\mathcal{U},\mathcal{X},f).\label{eq:rateRelation}
\end{align}
In Sec.~\ref{sec:converse}, we show the converse, namely that $S(P_Y)\geq\bar{S}(P_Y)\geq\entop(P_Y)$. In Sec.~\ref{sec:f2v}, we propose a fixed-to-variable length code and we show in Sec.~\ref{sec:proof} that it achieves in the limit a resolution rate of $\entop(P_Y)$. Combining converse and achievability proves the following proposition.
\begin{proposition}\label{prop:resolvability}
For any target distribution $P_Y$, the resolvability $S(P_Y)$ and the mean-resolvability $\bar{S}(P_Y)$ are both equal to the entropy of the target distribution, i.e.,
\begin{align*}
S(P_Y)=\bar{S}(P_Y)=\entop(P_Y).
\end{align*}
\end{proposition}

\subsection{Relation to Han-Verdu Resolvability\cite{han1993approximation}}
\label{sec:hv}
The \emph{Han-Verdu Resolution Rate} $R_\mathrm{hv}$ of a variable length encoder is \cite[Sec.~II]{han1993approximation} 
\begin{equation*}
R_\mathrm{hv} := \frac{\log_2 M_X}{\expop[\ell(X)]}
\end{equation*}
where $M_X$ is the minimum integer for which $P_X$ is $M$-type. We have replaced $n$ in \cite[Eq.~(2.4)]{han1993approximation} by $\expop[\ell(X)]$ to account for the variable length codewords. For the target distribution $P_Y$, we define the \emph{achievable} Han-Verdu resolution rate in the same way as we defined the achievable resolution rate in Sec.~\ref{sec:def}. Analogously, we define the Han-Verdu \emph{resolvability} and we denote it by $S_\mathrm{hv}(P_Y)$. Achievability has been defined in \cite[Sec.~II]{han1993approximation} for variational distance. Here, we deliberately define the Han-Verdu quantities with respect to un-normalized informational divergence to be able to compare them to the quantities we defined in Sec.~\ref{sec:def}.

Recall that $\lmax$ is the length of the longest binary string in $\mathcal{U}$. The distribution $P_U$ is then $2^\lmax$-type. Since $X=f(U)$ where $f$ is a deterministic mapping, we have $1 \leq M_X \leq 2^\lmax$. Therefore,
\begin{align*}
R_\mathrm{hv} &= \frac{\log_2 M_X}{\expop[\ell(X)]} \\
    &\leq \frac{\lmax}{\expop[\ell(X)]}.
\end{align*}
Also, we have
\begin{align*}
R &= \frac{\expop[\ell(U)]}{\expop[\ell(X)]} \\
  &\leq \frac{\lmax}{\expop[\ell(X)]}.
\end{align*}
For variable length input encoders, we cannot establish a relation between $R_\mathrm{hv}$ and $R$ in general. For example, an encoder can map a certain binary input string $u\in\mathcal{U}$ of length $\lmax$ to a codeword $x\in\mathcal{X}$ and map no other binary string in $\mathcal{U}$ to $x$. For this encoder, $P_X$ will also be of $2^\lmax$-type and hence $R_\mathrm{hv} \geq R$. On the other hand, if the encoder maps all binary strings $u\in\mathcal{U}$ with $|\mathcal{U}|>1$ to a single $x\in\mathcal{X}$, then $P_X$ is a $1$-type distribution, $R_\mathrm{hv} = 0$ and hence $R > R_\mathrm{hv}$.

However, for fixed-to-variable length encoders, $\expop[\ell(U)] = \lmax$ because of the fixed length parsing dictionary $\mathcal{U}$ and therefore \begin{align*}
R= \frac{\expop[\ell(U)]}{\expop[\ell(X)]}  = \frac{\lmax}{\expop[\ell(X)]} \geq R_\mathrm{hv}.
\end{align*}
By \cite[Lemma~2]{han1993approximation}, $\entop(P_X) \leq \log_2 M_X$ for any distribution $P_X$ and therefore $R_\mathrm{hv} \geq \bar{R}$.
\begin{remark}
Since $R_\mathrm{hv} \geq \bar{R}$ as well as $R \geq \bar{R}$ for any variable length encoder, establishing the converse $\bar{R}\geq\entop(P_Y)$ directly provides a converse for $R_\mathrm{hv}$ and $R$ as well. Moreover, we will construct a fixed-to-variable length encoder that achieves $R=\entop(P_Y)$. Since for fixed-to-variable length encoders $R \geq R_\mathrm{hv} \geq \bar{R}$, this establishes the achievability for Han-Verdu resolvability (and mean-resolvability) as well, i.e.,
\begin{equation*}
S_\mathrm{hv}(P_Y) = \entop(P_Y).
\end{equation*}
\end{remark}

\section{Converse}
\label{sec:converse}
For any variable length encoder, the resolution rate is greater than or equal to the entropy rate, see \eqref{eq:rateRelation}. To get a general converse, we therefore lower bound achievable entropy rates. Furthermore, since $\expop[\ell(X)]\geq 1$, the normalized informational divergence is smaller or equal to the un-normalized informational divergence, i.e.,
\begin{align}
\frac{\kl(P_X\Vert P_Y^\mathcal{X})}{\expop[\ell(X)]}\leq\kl(P_X\Vert P_Y^\mathcal{X}).\label{eq:normalized}
\end{align}
If a code can drive the un-normalized informational divergence to zero, it automatically also drives the normalized informational divergence to zero. We therefore lower bound the entropy rates that are achievable with respect to the normalized informational divergence. The following proposition relates entropy rate and informational divergence. 
\begin{proposition}\label{prop:klen}
\begin{align}
\frac{\kl(P_X\Vert P_Y^\mathcal{X})}{\expop[\ell(X)]}\to 0\Rightarrow\left|\frac{\entop(P_X)}{\expop[\ell(X)]}-\entop(P_Y)\right|\to 0.
\end{align}
\end{proposition}
\begin{IEEEproof}
We give the proof in \cite{bocherer2013rooted}.
\end{IEEEproof}
We will use Prop.~\ref{prop:klen} both in the proof of the converse and in the proof of achievability.
\begin{proposition}\label{prop:converse}
If $\tilde{R}$ is for a target distribution $P_Y$ an achievable entropy rate with respect to the normalized informational divergence, then
\begin{align}
\tilde{R}=\entop(P_Y).\label{eq:converse1}
\end{align}
In particular,
\begin{align}
S(P_Y)\geq\bar{S}(P_Y)\geq\entop(P_Y)\label{eq:converse2}
\end{align}
where definitions of $S(P_Y)$ and $\bar{S}(P_Y)$ in \ref{sec:def} are based on un-normalized informational divergence.
\end{proposition}
\begin{IEEEproof}
The statement \eqref{eq:converse1} of the proposition follows from Prop.~\ref{prop:klen}. The statement \eqref{eq:converse2} follows from \eqref{eq:rateRelation}, \eqref{eq:normalized}, and statement \eqref{eq:converse1}.
\end{IEEEproof}

\section{Fixed-to-Variable Length Resolution Code}
\label{sec:f2v}

We construct a fixed-to-variable length code with $m$ input bits as follows. First, we construct a complete variable length codebook $\mathcal{X}$ with $|\mathcal{X}|=N=2^n$ codewords by applying Tunstall coding \cite[Sec.~2.10]{kramer2012information} to $P_Y$. Note that $n$ does not necessarily need to be an integer. According to the Tunstall Lemma \cite[p.~47]{masseyapplied1}, the distribution $P_Y^\mathcal{X}$ is close to uniform in the sense that the smallest and the greatest probabilities differ at most by a factor of
\begin{align}
\mu_Y:=\min_{a\in\supp P_Y}P_Y(a).
\end{align}
Note that $\mu_Y$ does only depend on $P_Y$ and is independent of $\setx$. In particular, the probability $P_Y^\mathcal{X}(x)$ is lower bounded as
\begin{align}
P_Y^\mathcal{X}(x)\geq \mu_Y2^{-n}\label{eq:pylower}
\end{align}
and it is upper bounded as
\begin{align}
P_Y^\mathcal{X}(x)\leq \frac{2^{-n}}{\mu_Y}.\label{eq:pyupper}
\end{align}
Next, we quantize $P_Y^\mathcal{X}$ using \cite[Alg.~1]{bocherer2013optimala} to obtain the $2^{m}$-type $P_X$ with the property
\begin{align}
P_X(a)\leq P_Y^\mathcal{X}(a)+2^{-m},\qquad\forall a\in\mathcal{X}.\label{eq:quantBounds}
\end{align}
The distribution $P_X$ can be generated by a many-to-one mapping $f\colon\{0,1\}^m\to\mathcal{X}$. The codebook is of size $2^n$ and the input of the encoder is of length $m$ bits. We define $q=m-n$. If $q>0$, the mapping $f$ is many-to-one (but possibly not onto), and if $q<0$, the mapping is not onto (but possibly many-to-one).
\subsection{Discussion of Code Construction}
\begin{itemize}
\item The Tunstall code guarantees that for each codeword $x\in\setx$, the target probability $P_Y^\setx(x)$ deviates from the uniform probability $1/2^n$ at most by a factor of $\mu_Y$, which is independent of $\setx$, $n$, and $m$. 
\item Because the input is of fixed length $m$, it is uniformly distributed over the dictionary $\setu=\{0,1\}^m$. A many-to-one mapping from $\setu$ to $\setx$ resolves the remaining difference between the generated uniform distribution over $\setu$ and the almost uniform target distribution over $\setx$.
\end{itemize}
\section{Achievability}
\label{sec:proof}

The following proposition states that the fixed-to-variable length resolution code that we defined in Sec.~\ref{sec:f2v} achieves in the limit the resolvability of the target distribution.
\begin{proposition}[Achievability]\label{prop:achievability}\ 
\begin{enumerate}
\item If $q\to\infty$, then
\begin{align}
\kl(P_X\Vert P_Y^\mathcal{X})\to 0.
\end{align}
\item If $q\to\infty$ and $q/n\to 0$, then the resolution rate approaches the resolvability of the target distribution, i.e.,
\begin{align}
R\to\entop(P_Y).
\end{align}
\end{enumerate}
\end{proposition}
We now prove both statements of the proposition.

\subsection{Informational Divergence}

We bound
\begin{align*}
D(P_X\Vert P_Y^\mathcal{X})&=\sum_{a\in\supp P_X}P_X(a)\log_2\frac{P_X(a)}{P_Y^\mathcal{X}(a)}\\
&\overleq{a}\sum_{a\in\supp P_X}P_X(a)\log_2\frac{P_Y^\mathcal{X}(a)+2^{-m}}{P_Y^\mathcal{X}(a)}\\
&\overleq{b}\sum_{a\in\supp P_X}P_X(a)\log_2\left(1+\frac{2^{n}}{2^m\mu_Y}\right)\\
&\overleq{c}\sum_{a\in\supp P_X}P_X(a)\frac{2^{n}}{2^m\mu_Y}\log_2e\\
&=\frac{2^{n}}{2^m\mu_Y}\log_2e\\
&=\frac{2^{-q}}{\mu_Y}\log_2e\numberthis \label{eq:divbound}
\end{align*}
where (a) follows from \eqref{eq:quantBounds}, where (b) follows from \eqref{eq:pylower}, and where we used the bound $\log_2(1+x)\leq x\log_2 e$ in (c). The bound \eqref{eq:divbound} establishes the first statement of Prop.~\ref{prop:achievability}.

\subsection{Entropy}

For each $a\in\mathcal{X}$, the probability $P_X(a)$ is upper bounded by
\begin{align*}
P_X(a)&\overleq{a} P_Y^\mathcal{X}(a)+2^{-m}\\
&\overleq{b}\frac{2^{-n}}{\mu_Y}+2^{-m}\numberthis\label{eq:pxupper}
\end{align*}
where (a) and (b) follow from \eqref{eq:quantBounds} and \eqref{eq:pyupper}, respectively. We can now bound the entropy of $X$ as follows.
\begin{align*}
\entop(P_X)&=\expop\left[\log_2\frac{1}{P_X(X)}\right]\\
&\geq\log_2\frac{1}{\displaystyle\max_{a\in\supp P_X}P_X(a)}\\
&\overgeq{a} \log_2\frac{1}{\frac{2^{-n}}{\mu_Y}+2^{-m}}\\
&= n-\log_2\Bigl(\frac{1}{\mu_Y}+\frac{2^n}{2^m}\Bigr)\\
&= n-\log_2\Bigl(\frac{1}{\mu_Y}+2^{-q}\Bigr)\label{eq:hpxlower}\numberthis
\end{align*}
where we used \eqref{eq:pxupper} in (a). This can also be rewritten as a bound on $n$, namely
\begin{align}
n\leq \entop(P_X)+\log_2\left(\frac{1}{\mu_Y}+2^{-q}\right)\label{eq:nupper}.
\end{align}

\subsection{Rate}
For the rate, we get
\begin{align*}
R&=\frac{\expop[\ell(U)]}{\expop[\ell(X)]}\\  
&=\frac{m}{\expop[\ell(X)]}\\
&=\frac{n}{\expop[\ell(X)]}+\frac{q}{\expop[\ell(X)]}\\
&\overleq{a}\frac{\entop(P_X)}{\expop[\ell(X)]}+\frac{q+\log_2(\frac{1}{\mu_Y}+2^{-q})}{\expop[\ell(X)]}\numberthis \label{eq:ratebound}
\end{align*}
where we used \eqref{eq:nupper} in (a). We separately bound the two terms in \eqref{eq:ratebound}. For the second term, we get
\begin{align*}
\frac{q+\log_2(\frac{1}{\mu_Y}+2^{-q})}{\expop[\ell(X)]}&\overleq{a} \frac{q+\log_2(\frac{1}{\mu_Y}+2^{-q})}{\entop(P_X)\frac{1}{\log_2 D}}\\
&\overleq{b}  \frac{q+\log_2(\frac{1}{\mu_Y}+2^{-q})}{n-\log_2(\frac{1}{\mu_Y}+2^{-q})}\log_2 D\\
&\overset{\frac{q}{n}\to 0,q\to\infty}{\to}0 \numberthis \label{eq:ratebound1}
\end{align*}
where (a) follows from the Source Coding Theorem \cite[Theo~4.1]{csiszar2011information} and where we used \eqref{eq:hpxlower} in (b). For the first term in \eqref{eq:ratebound}, we have by \eqref{eq:divbound} and Proposition~\ref{prop:klen},
\begin{align}
\frac{\entop(P_X)}{\expop[\ell(X)]}\overset{q\to\infty}{\to}\entop(P_Y).\label{eq:ratebound2}
\end{align}
Using \eqref{eq:ratebound1} and \eqref{eq:ratebound2} in \eqref{eq:ratebound}, we get
\begin{align}
\lim_{\stackrel{\frac{q}{n}\to 0}{q\to\infty}}R\leq\entop(P_Y).\label{eq:rlim}
\end{align}
From the converse stated in Prop.~\ref{prop:converse}, we know that 
\begin{align}
 R \geq \entop(P_Y).\label{eq:rconverse}
\end{align}
Combining \eqref{eq:rlim} and \eqref{eq:rconverse}, we conclude that
\begin{align}
R\overset{\stackrel{\frac{q}{n}\to 0}{q\to\infty}}{\to}\entop(P_Y).
\end{align}
This proves the second statement of Prop.~\ref{prop:achievability}.

\section{Numerical Example}
\label{sec:numerical}
\begin{figure}
\centering
\footnotesize
\def\svgwidth{1.0\columnwidth}
\executeiffilenewer{figures/ratedivergence.svg}{figures/ratedivergence.pdf}%
{inkscape -z -D --file=figures/ratedivergence.svg %
--export-pdf=figures/ratedivergence.pdf --export-latex}%
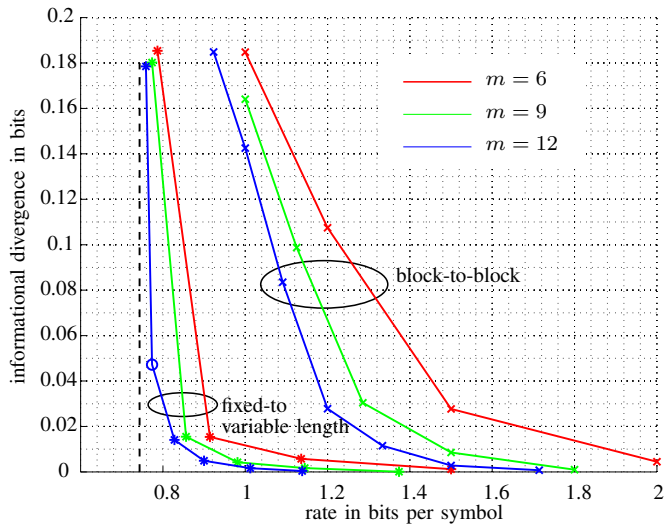%

\caption{Comparison of fixed-to-variable length and block-to-block resolution coding. The target distribution is $P_Y(0)=1-P_Y(1)=0.211$. The input length takes the values $m=6,9,12$ (red,green,blue, respectively). For each $m$, $n$ is varied, see Tab.~\ref{tab:mn}. For a fixed input length $m$, a larger $n$ results in a smaller rate but a higher divergence. The divergence-rate pairs of block-to-block coding are marked by an $\times$ and the divergence-rate pairs of fixed-to-variable length coding are marked by a $*$. The vertical dashed line marks the theoretical limit of $\entop(P_Y)=0.7415$. We use codebook sizes of $N=2^n$ for the fixed-to-variable length code for comparison. However, we are not restricted to that. To illustrate this, we consider $m=12$ and $2^n=N=3072$, which is not an integer power of two. The resulting divergence-rate pair is marked by a circle.}
\label{fig:example}
\end{figure}
To evaluate the performance of our resolution code, we fix $m$ and then determine, which rate-divergence pairs we can achieve with $m$ fair bits. Our example is the binary target distribution
\begin{align}
P_Y(0)=0.211,\quad P_Y(1)=0.789.
\end{align}
We compare our fixed-to-variable length resolution code from Sec.~\ref{sec:f2v} to optimal block-to-block resolution coding.

\subsection{Optimal Block-to-Block Resolution Coding}

Consider an $m$-to-$n$ block-to-block encoder. The codebook is $\mathcal{Y}^n$, i.e., the $n$-fold Cartesian product of $\mathcal{Y}$. Note that $n$ has to be an integer. The encoder defines a deterministic mapping from $\mathcal{U}=\{0,1\}^m$ to $\mathcal{Y}^n$. The distribution of $X=f(U)$ is $2^m$-type. The optimal encoder chooses the $f$ that generates the $2^m$-type quantization $P_X^*$ that minimizes the informational divergence, i.e.,
\begin{align}
P_X^*=\argmin_{\text{$P$ is $2^m$-type}}\kl(P\Vert P_Y^n).
\end{align}
The $2^m$-type $P_X^*$ can efficiently be found by \cite[Alg.~2]{bocherer2013optimala}. The resulting rate is $\frac{m}{n}$ and the resulting divergence is $\kl(P_X^*\Vert P_Y^n)$.

\subsection{Discussion}

We calculate the divergence-rate pairs that are achieved by block-to-block and variable-to-fixed length codes for several values of $m$ and $n$, see Tab.~\ref{tab:mn}. The results are displayed in Fig.~\ref{fig:example}. The theoretical minimum rate $\entop(P_Y)=0.7415$ is marked by a vertical dashed line. As we can see, for larger $m$, the divergence-rate curves get closer to the theoretical rate limit of $\entop(P_Y)$ and the divergence limit of $0$. Fixed-to-variable length coding significantly outperforms block-to-block coding. For $m=12$, the fixed-to-variable length curve is remarkably close to the limits. It should be remarked that the codebook size for fixed-to-variable length coding is not restricted to integer powers of two. We illustrate this by displaying a divergence-rate pair that is achieved by $2^n=N=3072$. It is marked by a circle in Fig.\ref{fig:example}.
\begin{table}
\centering
\caption{codebook sizes}
\label{tab:mn}
\begin{tabular}{r|ccccc}
m&\multicolumn{5}{c}{$n$}\\\hline
$6$&$3$&$4$&$5$&$6$&\\
$9$&$5$&$6$&$7$&$8$&$9$\\
$12$&$8$&$9$&${10}$&${11}$&${12}$
\end{tabular}
\end{table}

\section*{Acknowledgment}

The authors thank J. Hou and G. Kramer for useful discussions.

\bibliographystyle{IEEEtran}
\bibliography{IEEEabrv,confs-jrnls,references}

\end{document}